\documentclass[twocolumn,showpacs,preprintnumbers,amsmath,amssymb]{revtex4}

\usepackage{graphicx}
\usepackage{dcolumn}
\usepackage{bm}

\begin{document}

\preprint{APS/PRB}

\title{Incommensurate magnetic order in Ag$_{2}$NiO$_{2}$}

\author{J. Sugiyama$^1$}
 \email{e0589@mosk.tytlabs.co.jp}
\author{Y. Ikedo$^1$}
\author{K. Mukai$^1$}
\author{J. H. Brewer$^2$}
\author{E. J. Ansaldo$^3$}
\author{G. D. Morris$^3$}
\author{K. H. Chow$^{4}$}
\author{H. Yoshida$^5$}
\author{Z. Hiroi$^5$}
\affiliation{%
$^1$Toyota Central Research and Development Labs. Inc., 
 Nagakute, Aichi 480-1192, Japan}%

\affiliation{$^2$TRIUMF, CIAR and 
Department of Physics and Astronomy, University of British Columbia, 
Vancouver, BC, V6T 1Z1 Canada 
}%

\affiliation{$^3$TRIUMF, 4004 Wesbrook Mall, Vancouver, BC, V6T 2A3 Canada 
}%

\affiliation{$^4$Department of Physics, University of Alberta, Edmonton, AB, T6G 2J1 Canada
}%

\affiliation{$^5$ISSP, University of Tokyo, 
5-1-5 Kashiwanoha, Kashiwa 277-8581, Japan
}%

\date{\today}

\begin{abstract}
The nature of the magnetic transition of the
half-filled triangular antiferromagnet Ag$_{2}$NiO$_2$ with $T_{\rm N}$=56K was studied
with positive muon-spin-rotation and relaxation ($\mu^+$SR)
spectroscopy.
Zero field $\mu^+$SR measurements indicate
the existence of a static internal magnetic field at temperatures below $T_{\rm N}$.
Two components with slightly different precession frequencies and 
wide internal-field distributions suggest the formation 
of an incommensurate antiferromagnetic order below 56~K.
This implies that the antifrerromagnetic interaction is predominant in the NiO$_2$ plane 
in contrast to the case of the related compound NaNiO$_2$. 
An additional transition was found at $\sim$22~K by 
both $\mu^+$SR and susceptibility measurements. 
It was also clarified that the transition at $\sim$260~K 
observed in the susceptibility of Ag$_{2}$NiO$_{2}$
is induced by a purely structural
transition.
\end{abstract}


\maketitle

\section{\label{sec:Intro}Introduction}

Two-dimensional triangular lattice (${\sf 2DTL}$) antiferromagnets 
with a half-filled ($e_{g}$) state 
exhibit a variety of magnetically ordered states
due to competition between the
antiferromagnetic (AF) interaction and geometrical frustration. 
The discovery of superconductivity in Na$_{0.35}$CoO$_2\ast$1.3H$_2$O \cite{NCO_sc_1}
leads to an additional interest in the possible relationship 
between magnetic and superconducting order parameters 
in the ${\sf 2DTL}$ near half-filling.
The layered nickel dioxides,  
a series of materials with chemical formula $A^+$Ni$^{3+}$O$_2$, 
such as rhombohedral LiNiO$_2$,\cite{LNO_1,LNO_2},  
NaNiO$_2$\cite{NNO_1,NNO_2,NNO_3}, 
and AgNiO$_2$,\cite{ANO_1,ANO_2} 
in which Ni ions form the ${\sf 2DTL}$ by the connection of edge-sharing NiO$_6$ octahedra,
has been considered to be good candidates for an ideal half-filled ${\sf 2DTL}$.
In these materials at low temperature, there is a strong interaction
between the  
Ni$^{3+}$ ions and the crystalline electric field of the NiO$_{6}$ octahedron.
This causes the Ni$^{3+}$ ions to be in the 
low spin state with a $t_{2g}^6e_g^1$ ($S$=1/2) 
configuration.
 
Among the three layered nickel dioxides, 
NaNiO$_2$ is perhaps the best investigated.
It exhibits 
two transitions at  $T_{\rm JT}\sim$480~K and $T_{\rm N}$=23~K.
The former is a cooperative Jahn-Teller (JT) transition from 
a high-$T$ rhombohedral phase to a low-$T$ monoclinic phase, 
while the latter is a transition into an A-type AF phase ---
i.e. ferromagnetic (FM) order in the NiO$_2$ plane 
but AF between the two adjacent NiO$_2$ planes, 
as has been
reconfirmed very recently by both neutron diffraction\cite{NNO_1,NNO_2} and 
positive muon spin rotation/relaxation ($\mu^+$SR) experiments.\cite{NNO_3}  
The magnetic order is associated with the JT induced trigonal distortion
which stabilizes a half occupied $d_{z^2}$ orbital.\cite{NNO_4}   

Although
LiNiO$_2$ and NaNiO$_{2}$ are structurally very similar, 
LiNiO$_2$ shows dramatically different magnetic properties.
LiNiO$_2$ exhibits neither a cooperative JT transition
nor long-range magnetic order down to the lowest $T$ investigated. 
In fact, both heat capacity and NMR measurements suggest 
a spin-liquid state with short-range FM correlations.\cite{LNO_1}
Chatterji {\it et al.}\cite{LNO_2}, however,
found a rapid increase in the muon spin relaxation rates in LiNiO$_2$ 
below $\sim$10~K using the longitudinal field-$\mu^+$SR technique,
suggesting a spin-glass-like behavior below 10~K. 
The discrepancy between the two results is considered to be 
a sample-dependent phenomenon that arises from  
the difficulties in preparing stoichiometric LiNiO$_2$.
The third compound, AgNiO$_2$, also lacks a cooperative JT transition.
A magnetic transition
$T_{\rm N}$ was clearly observed by 
both susceptibility ($\chi$) and $\mu^+$SR measurements
but long-range magnetic order was not detected 
by a neutron diffraction experiment even at 1.4~K.\cite{ANO_2}

While the nature of the 
magnetic ground states of LiNiO$_2$ and AgNiO$_2$ is still not clear, 
the FM interaction on the ${\sf 2DTL}$ NiO$_2$ plane has been 
thought to be common for all the layered Ni dioxides with a half-filled state 
because of the clear magnetic order observed in NaNiO$_2$.
In this paper, we present measurements that demonstrate
this supposition is incorrect. 
This is accomplished by investigating the magnetism in Ag$_{2}$NiO$_{2}$,
a material that can be represented by the chemical formula 
(Ag$_2$)$^+$Ni$^{3+}$O$_2$ and hence is expected to have 
a NiO$_{2}$ plane
that has properties identical to the above three
layered nickel dioxides.
However, in Ag$_{2}$NiO$_{2}$, static AF order,
likely the formation of an incommensurate AF structure 
in the NiO$_2$ plane, is observed instead.
 
Disilver nickel dioxide Ag$_2$NiO$_2$ 
is a rhombohedral system with space group $R\overline{3}m$ 
($a_{\rm H}=0.29193$~nm and $c_{\rm H}=2.4031$~nm 
for the hexagonal unit-cell) \cite{A2NO_1}
that was found to exhibit two transitions at $T_{\rm S}$=260~K and $T_{\rm N}$=56~K 
by resistivity and $\chi$ measurements, 
while the symmetry remains rhombohedral down to 5~K.\cite{A2NO_3} 
Interestingly, Ag$_2$NiO$_2$ shows metallic conductivity down to 2~K 
probably due to a quarter-filled Ag 5$s$ band,
as in the case of Ag$_2$F.\cite{A2F_1}
Very recently, Yoshida {\it et al.} proposed the significance of 
the AF interaction in the ${\sf 2DTL}$ NiO$_2$ plane 
from the $\chi(T)$ measurement.\cite{A2NO_3}
 

\section{\label{sec:Exp}Experimental}

A powder sample of Ag$_2$NiO$_2$ was prepared  
at the ISSP of the University of Tokyo 
by a solid-state reaction technique 
using reagent grade Ag$_2$O  
and NiO powders as starting materials.
A mixture of Ag$_2$O and NiO 
was heated at 550$^{\rm o}$C for 24~h in oxygen under a
pressure of 70~MPa.
A more detailed description of the 
preparation and characterization of the powder 
is presented in Ref.~\onlinecite{A2NO_3}. 

Susceptibility ($\chi$) was measured 
using a superconducting quantum interference device 
({\sf SQUID}) magnetometer 
(mpms, Quantum Design) 
in the temperature range between 400 and 5~K 
under magnetic field $H \leq$ 55~kOe. 
For the $\mu^+$SR experiments, 
the powder was pressed into a disk of about 20~mm diameter and thickness 1~mm, 
and subsequently placed in a muon-veto sample holder. 
The $\mu^+$SR spectra were measured on 
the {\sf M20} surface muon beam line at TRIUMF. 
The experimental setup and techniques
were described elsewhere.\cite{SDW_2} 

\section{\label{sec:Re}Results and Discussion}
\subsection{Below $T_{\rm N}$}

\begin{figure}[h]
\includegraphics[width=7.cm]{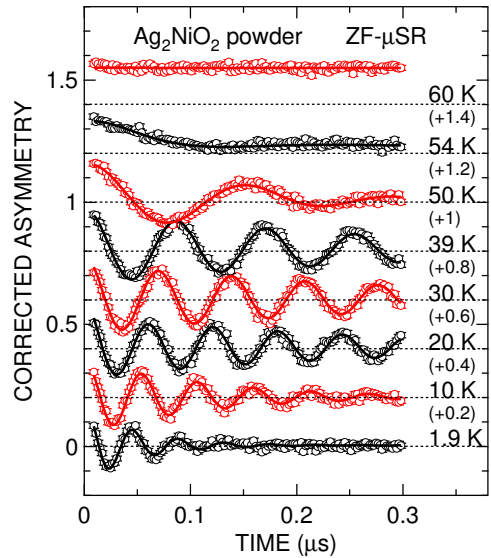}
\caption{\label{fig:ZF-muSR1} (Color online) 
Temperature dependence of the ZF-$\mu^+$SR time spectra of 
  a powder sample of Ag$_{2}$NiO$_2$.
  Each spectrum is offset by 0.2 for clarity of the display. 
  The solid lines represent the fitting result using Eq.~(\ref{eq:ZFfit}).
}
\end{figure}
\begin{figure}[h]
\includegraphics[width=7.cm]{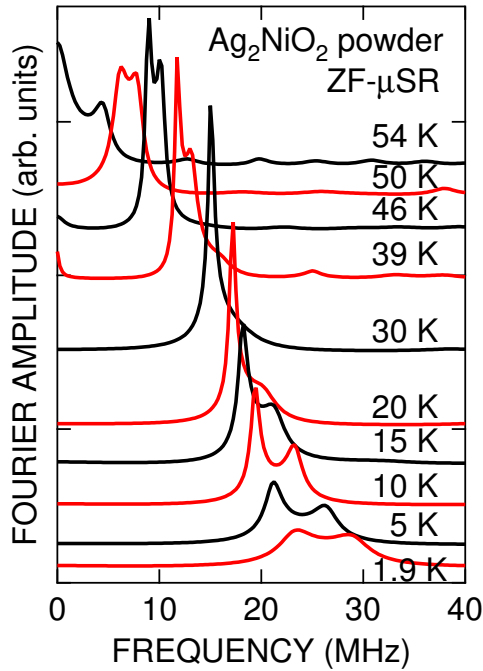}
\caption{\label{fig:FFT} (Color online) 
Temperature dependence of the 
  Fourier Transform of the ZF-$\mu^+$SR time spectrum 
  for Ag$_{2}$NiO$_2$.   
}
\end{figure}
\begin{figure}[h]
\includegraphics[width=6.cm]{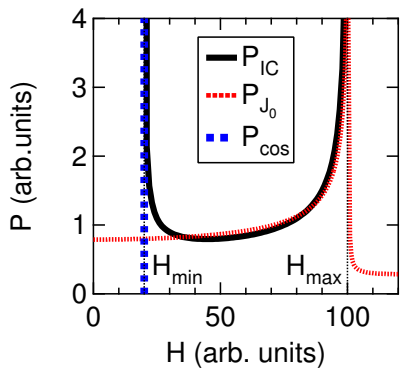}
\caption{\label{fig:ICSDW} (Color online) 
The distribution of the 
magnitude of the magnetic field $H$ due to a 
generic incommensurate magnetic structure 
described in the text.
The distribution corresponding to a Bessel function $J_0(\omega_2 t)$ 
and a $\cos(\omega_1 t)$ are also plotted for comparison. 
Here, $\gamma_{\mu}H_{\rm max}=\omega_2$ and 
$\gamma_{\mu}H_{\rm min}=\omega_1$.
}
\end{figure}

Figure~\ref{fig:ZF-muSR1} shows zero-field (ZF-)$\mu^+$SR time spectra 
in the $T$ range between 1.9~K and 60~K  
for a powder sample of Ag$_{2}$NiO$_2$. 
A clear oscillation due to quasi-static internal fields ${\bm H}_{\rm int}$ 
is observed below 54~K, unambiguously establishing
the existence  of long-range magnetic order in the sample.
Interestingly, as $T$ is decreased from 60~K, 
the relaxation rate first decreases down to $\sim$20~K 
and then {\it increases} as $T$ is lowered further.
By contrast, the average
oscillation frequency increases monotonically down to 1.9~K. 
This implies that  
the distribution of ${\bm H}_{\rm int}$ at $T\geq$ 54~K and $\leq$ 20~K 
is larger than that at 20~K$<T<$54~K. 

This is further established by the 
$T$ dependence of the Fourier Transform of 
the ZF-$\mu^+$SR time spectrum shown in  Fig.~\ref{fig:FFT}.
Note that there is clearly line broadening
below 20~K as well as above 54~K. 
The line-broadening above 54~K is reasonably explained 
by critical phenomena in the vicinity of $T_{\rm N}$=56~K;
however, it is difficult to understand the origin of the line-broadening below 20~K 
using a classical AF model without invoking
the presence of an additional magnetic transition. 
Furthermore, even the spectrum at 30~K, which is the sharpest FFT measured, 
consists of a main peak at $\sim$14~MHz and a shoulder around 16~MHz, 
suggesting a wide distribution of ${\bm H}_{\rm int}$ in Ag$_{2}$NiO$_2$.

We therefore use a combination of three signals to fit  
the ZF-$\mu^+$SR time spectrum: 
\begin{eqnarray}
 A_0 \, P_{\rm ZF}(t) &=& 
   A_{1} \, \cos(\omega_{\mu,1} t + \phi) \, \exp(-\lambda_{1} t)
\cr
 &+& A_{2} \, J_0(\omega_{\mu,2} t) \, \exp(-\lambda_{2} t)
\cr
 &+& A_{\rm slow} \, \exp(-\lambda_{\rm slow} t),
\label{eq:ZFfit}
\end{eqnarray}
where $A_0$ is the empirical maximum muon decay asymmetry, 
$A_{1}$, $A_{2}$ and $A_{\rm slow}$ 
are the asymmetries associated with the three signals,  
$J_0(\omega_{\mu,2} t)$ is a zeroth-order Bessel function of the first kind  
that describes the muon polarization evolution 
in an incommensurate spin density wave (IC-SDW) field distribution,\cite{SDW_2} 
and $\omega_{\mu,1}<\omega_{\mu,2}$. 

Although $J_0(\omega t)$ is widely used for fitting the ZF-$\mu^+$SR spectrum 
in an IC-SDW state, 
it should be noted that $J_0(\omega t)$ only provides 
an approximation of the generic IC magnetic field distribution. 
This is because the lattice sum calculation
of the dipole field at the muon site (${\bm H_{\rm IC}}$)
due to an IC magnetic structure lies in a plane and traces
out an ellipse.
The half length of the major axis of the ellipse corresponds to $H_{\rm max}$, 
whereas half of the minor axis corresponds to $H_{\rm min}$.
As a result, the IC magnetic field distribution $P_{\rm IC}$ is generally given by;
\begin{eqnarray}
 P_{\rm IC}=P({\bm H_{\rm IC}})=\frac{2}{\pi}\frac{H}{\sqrt{(H^2-H_{\rm min}^2)(H_{\rm max}^2-H^2)}}.
\label{eq:PIC}
\end{eqnarray}
The distribution diverges as $H$ 
approaches either $H_{\rm min}$ or $H_{\rm max}$ 
(see Fig.~\ref{fig:ICSDW}). 
$J_0(\omega t)$ describes the field distribution
very well except in the vicinity of $H_{\rm min}$,
and the value of $\omega$ should be interpreted as an accurate
measure of  $H_{\rm max}$.
However, $J_0(\omega t)$ provides no information 
on $H_{\rm min}$. 
Hence, 
the first term $A_1\cos(\omega_{\mu,1}t+\phi_1)\exp(-\lambda_1t)$ is 
added in Eq.~(\ref{eq:ZFfit}) to 
account for the intensity around $H_{\rm min}$ 
and to determine the value of 
$H_{\rm min}$(= $\omega_{\mu,1}/\gamma_{\mu}$) \cite{Andreica_1}
($\gamma_{\mu}$ is the muon gyromagnetic ratio and  
$\gamma_{\mu}/2\pi$=13.55342~kHz/Oe).
In other words, only when $H_{\rm min}$=0, 
Eq.~(\ref{eq:PIC}) is well approximated by $J_0(\omega t)$. 
Here it should be emphasized that 
$\mu^+$SR spectra are often fitted 
in a time domain, 
i.e. not by Eq.~(\ref{eq:PIC}) but by Eq.~(\ref{eq:ZFfit}), 
since information on all the parameters 
such as $A$, $\omega$, $\lambda$ and $\phi$ are necessary 
to discuss the magnetic nature of the sample.  

We note that the data can also be
well-described using two cosine oscillation signals, 
$A_1\cos(\omega_{\mu,1}t+\phi_1)\exp(-\lambda_1t) 
+ A_2\cos(\omega_{\mu,2}t+\phi_2)\exp(-\lambda_2t)$ with
$\phi_2$=-54$\pm$10$^{\rm o}$ below $T_{\rm N}$. 
The delay is physically meaningless,
implying that the field distribution fitted by a cosine oscillation,
i.e. a commensurate ${\bm H_{\rm int}}$ does not exist in Ag$_2$NiO$_2$.\cite{SDW_2}
Furthermore, as $T$ decreases from 54~K, 
$A_1$ ($A_2$) decreases (increases) linearly with $T$ 
from 0.15 (0) at 54~K to 0 (0.15) at 1.9~K. 
In order to explain the $A_1(T)$ and $A_2(T)$ curves,
one would need to invoke the existence of two muon sites, 
and a situation whereby the 
population of $\mu^+$ at each site 
is changing in proportion to $T$.
Such behavior is very unlikely to occur at low $T$.  
Hence, we believe that our data strongly
suggests the appearance of an IC-AF order in Ag$_2$NiO$_2$ 
below $T_{\rm N}$, 
as predicted by the calculation using a Mott-Hubbard model (discussed later). 
Such a conclusion is also consistent with the fact that 
the paramagnetic Curie temperature is -33~K 
estimated from the $\chi(T)$ curve below 260~K.\cite{A2NO_3} 

\begin{figure}
\includegraphics[width=6.cm]{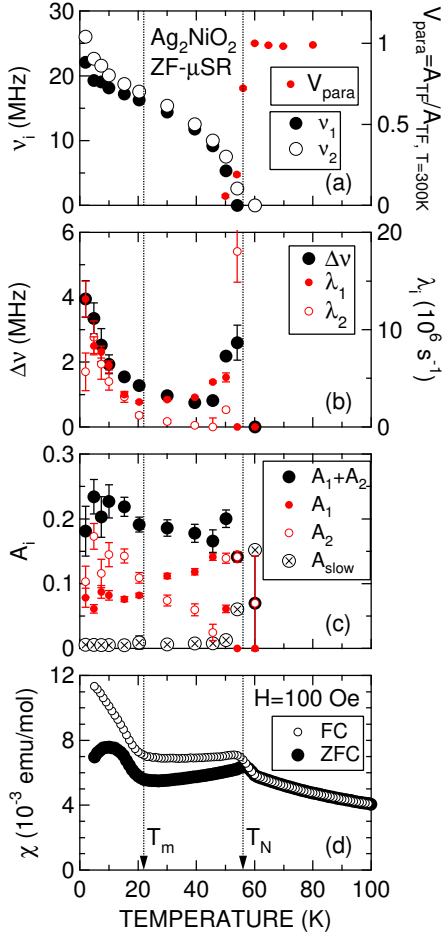}
\caption{\label{fig:ZF-muSR} (Color online) 
Temperature dependences of  
(a) the muon precession frequencies ($\nu_ i$=$\omega_{\mu, i}/2\pi$) 
and normalized transverse field asymmetry that roughly corresponds to 
the volume fraction of the paramagnetic phases in the sample ($V_{\rm para}$), 
(b) $\Delta\nu =\nu_2-\nu_1$, $\lambda_1$ and $\lambda_2$,
(c) the asymmetries $A_1$+$A_2$, $A_1$, $A_2$ and $A_{\rm slow}$   
and (d) $\chi$ 
for the powder sample of Ag$_{2}$NiO$_2$. 
$\chi$ was measured in zero-field-cooling $ZFC$ and field-cooling $FC$ mode
with $H=100$~Oe. 
}
\end{figure}
Figures~\ref{fig:ZF-muSR}(a) - \ref{fig:ZF-muSR}(d) show 
the $T$ dependence of  
the muon precession frequencies ($\nu_i$=$\omega_{\mu, i}/2\pi$), 
the volume fraction of the paramagnetic phases ($V_{\rm para}$), 
$\Delta\nu =\nu_2-\nu_1$, $\lambda_1$, $\lambda_2$,
the asymmetries $A_1$+$A_2$, $A_1$, $A_2$, $A_{\rm slow}$,   
and $\chi$ 
for the powder sample of Ag$_{2}$NiO$_2$. 
Here, $V_{\rm para}$ is estimated from the weak transverse field (wTF-) 
$\mu^+$SR experiment described later.  
In agreement with the FFTs shown in Fig.~\ref{fig:ICSDW}, 
as $T$ is decreased from 60~K, 
$\nu_2$ appears at 54~K.
It then increases monotonically with decreasing $T$ down to around 20~K, 
and then increases more rapidly upon further cooling.
The $\nu_1(T)$ curve exhibits a similar behavior 
to that observed for $\nu_2(T)$.
It is noteworthy that 
as $T$ is decreased from 80~K, 
the $V_{\rm para}(T)$ curve shows a sudden drop down to $\sim0$ at $T_{\rm N}$, 
indicating that the whole sample enters into an IC-AF state. 

As $T$ decreases from $T_{\rm N}$, 
$\Delta\nu$, 
which measures the distribution of ${\bm H}_{\rm int}$ in the IC-AF phase, 
rapidly decreases down to $\sim$0.8~MHz at 40~K, 
then seems to level off the lowest value down to $\sim$20~K
and then increases with increasing slope ($\mid$d$\Delta\nu$/d$T\mid$) 
until it reaches 4~MHz at 1.9~K. 
The overall $T$ dependence of $\Delta\nu$ is similar to that of $\lambda_i$.
This behavior is expected since 
a large $\Delta\nu$ naturally
implies a more inhomogeneous field distribution--- 
i.e, an increased flattening of the ellipse
that enhances $\lambda_i$. 
The asymmetry of the IC magnetic phase, $A_1$+$A_2$,
also increases monotonically 
with decreasing $T$, 
although a small jump likely exists around 20~K.
The existence of a significant $A_1$ underscores the 
inappropriateness of fitting the ZF-$\mu^+$SR
data with only a $J_0(\omega_{\mu,2} t)$ term.
In fact, note that $A_{1} < A_{2}$ above 20 K,
suggesting that the IC-AF order develops/completes
below 20 K. 
This is consistent with the rapid increases in $\Delta\nu$ and $\lambda_i$
below 20~K, as described above. 

The behavior of the muon parameters
is quite consistent with the $\chi(T)$ curve, 
which exhibits a sudden increase in the slope ($\mid$d$\chi_{\rm FC}$/d$T\mid$) 
below $\sim$22~K(=$T_{\rm m}$)
with decreasing $T$. 
Note the $\chi(T)$ curve measured under
$ZFC$ conditions starts to deviate 
from that measured in the $FC$ configuration below $T_{\rm N}$, 
suggesting the development of a ferro- or ferrimagnetic component 
probably due to a canted spin structure.\cite{NNO_2}
\begin{figure}[t]
\includegraphics[width=6.cm]{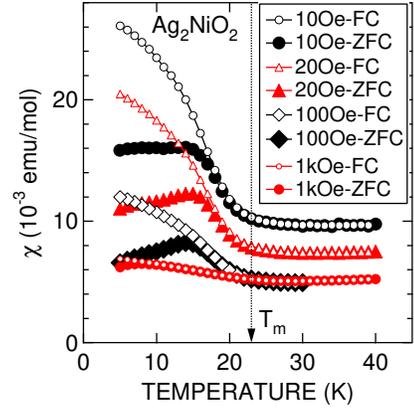}
\caption{\label{fig:chi} (Color online) 
Temperature dependence 
of $\chi$ measured in both $ZFC$ and $FC$ mode 
well below $T_{\rm N}=56~$K 
with $H$=10~Oe, 20~Oe, 100~Oe and 1~kOe for Ag$_{2}$NiO$_2$. 
}
\end{figure}
The ferro- or ferrimagnetic behavior is however observed only 
at low $H$, 
although the cusp at $T_{\rm N}$ is clearly seen with $H$=100~-~10~kOe 
(see Figs.~\ref{fig:ZF-muSR}(d) and ~\ref{fig:wTF}(d)). 
Below $T_{\rm m}$, $\chi_{\rm FC}$ increases with decreasing $T$, 
while the slope is suppressed by increasing $H$ (see Fig.~\ref{fig:chi}).
Keeping in mind that $\mu^+$SR is insensitive to magnetic impurities, 
we conclude that Ag$_{2}$NiO$_2$ undergoes 
a transition from a paramagnetic to 
an IC-AF state at $T_{\rm N}$=56~K
and then to a slightly different ordered state at $T_{\rm m}\sim$22~K. 

It is worth contrasting the current 
$\mu^+$SR results on Ag$_2$NiO$_2$ with those 
in related compounds NaNiO$_2$ and AgNiO$_2$.
The ZF-$\mu^+$SR spectrum on a powder sample of NaNiO$_2$ 
consists of two signals below $T_{\rm N}(\sim$20~K): 
an exponentially relaxing cosine oscillating signal 
(same as the first term in Eq.~(\ref{eq:ZFfit}))
as the predominant component and 
a minor signal described by an exponential relaxation.\cite{NNO_3} 
This indicates that the whole NaNiO$_2$ sample enters into 
a commensurate AF state below $T_{\rm N}$, 
being consistent with the magnetic structure determined by neutron diffraction experiments, 
i.e., an A-type AF order.\cite{NNO_1,NNO_2}  
Interestingly, the value of $\nu_{\rm T\rightarrow0~K}$=64.2~MHz, 
which corresponds to $H_{\rm int}\sim$0.5~T, 
is 2.5 times higher than that for Ag$_{2}$NiO$_2$. 
The muon site in NaNiO$_2$ is assigned to the vicinity of the O ions,\cite{NNO_3} and 
is thought to be also reasonable for the other layered nickel dioxides. 
The differences between the $\mu^+$SR results on NaNiO$_2$ and Ag$_2$NiO$_2$ 
hence suggest that the magnetic structure of Ag$_{2}$NiO$_2$ 
is most unlikely to be an A-type AF.
Furthermore, there are no indications for additional transitions of NaNiO$_2$
below $T_{\rm N}$ 
by $\chi$, $\mu^+$SR and neutron diffraction measurements.\cite{NNO_1,NNO_2,NNO_3}

In AgNiO$_2$, the primary ZF-$\mu^+$SR signal
is one that exponentially relaxes down to the lowest $T$
($\sim$3~K).
Below $T_{\rm N}$(=28~K), three minor oscillating components appear. 
These have small amplitudes and correspond to internal fields from 
0.2 to 0.33~T (27~-~45~MHz).\cite{ANO_2} 
The comparison between AgNiO$_2$ and Ag$_2$NiO$_2$ indicates 
that the interlayer separation ($d_{\rm NiO_2}$) enhances 
the static magnetic order in the NiO$_2$ plane. 
It is highly unlikely that 
the AF interaction through the double Ag$_2$ layer 
is stronger than that through the single Ag layer, 
since $d_{\rm NiO_2}$=0.801~nm for Ag$_2$NiO$_2$\cite{A2NO_1} and 
0.612~nm for AgNiO$_2$.\cite{ANO_1}

Our results therefore suggest
that the AF order exists in the NiO$_2$ plane, 
in contrast to the situation in NaNiO$_2$. 
Assuming the AF interaction is in the NiO$_2$ plane, 
an IC- spiral SDW phase 
is theoretically predicted to appear in a 
half-filled {\sf 2DTL}\cite{2DTL_1}  
(calculated using  the Hubbard model within a mean field 
approximation with $U/t\geq3.97$, 
where $U$ is the Hubbard on-site repulsion and 
$t$ is the nearest-neighbor hopping amplitude). 
In order to further establish the magnetism in Ag$_{2}$NiO$_2$,
it would be interesting to carry out neutron diffraction experiments 
to determine the magnetic structure below $T_{\rm N}$ and below $T_{\rm m}$. 

We wish here to mention that 
if the valence state of the Ni ion in the NiO$_2$ plane 
can be varied for Ag$_2$NiO$_2$, 
the resultant phase diagram should serve as an interesting
comparison with 
that of  $A_x$CoO$_2$ ($A$=alkali elements) with $x\leq$0.5. 
Unlike Li$_x$NiO$_2$, 
(Ag$_2$)-deficient samples are currently unavailable, 
probably because of the metal-like Ag-Ag bond in the disilver layer.\cite{A2NO_1}  
A partial substitution for Ag$_2$ by other cations
has thus far also been unsuccessful for reasons unknown.  

\subsection{Near $T_{\rm S}$}
\begin{figure}[h]
\includegraphics[width=6.cm]{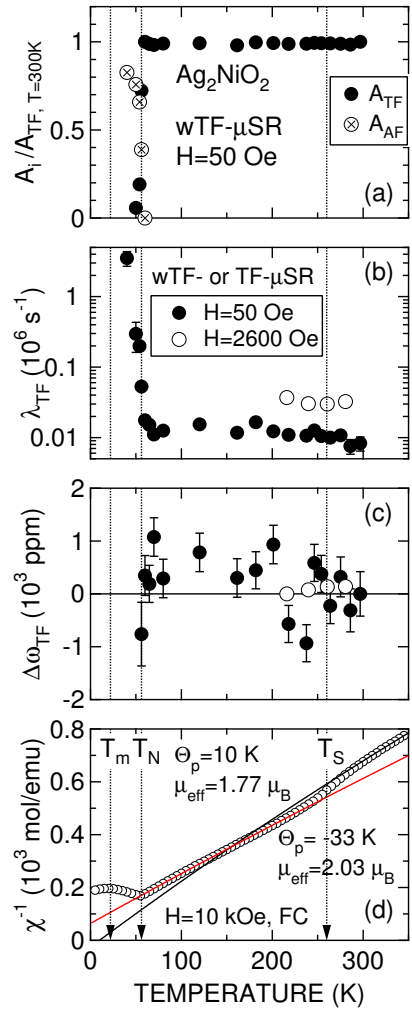}
\caption{\label{fig:wTF} (Color online) 
Temperature dependences of 
(a) the normalized $A_{\rm TF}$ and $A_{\rm AF}$, 
(b) $\lambda_{\rm TF}$,  
(c) the shift of the muon precession frequency, $\Delta\omega_{\mu, \rm TF}$ and 
(d) the inverse susceptibility, $\chi^{-1}$ 
in the Ag$_{2}$NiO$_2$ powder. 
The wTF and TF data were obtained by fitting using Eq.~(\ref{eq:TFfit}). 
$\chi$ was measured in $FC$ mode with $H$=10~kOe.
The paramagnetic Curie temperature ($\Theta_{\rm p}$) and 
the effective magnetic moment of Ni ions ($\mu _{\rm eff}$) 
are calculated above and below $T_{\rm S}$ 
by the Curie-Weiss law in the general form; 
$\chi = C(T-\Theta _{\rm p})^{-1}~+~\chi _0$.
}
\end{figure}
In order to elucidate the magnetic behavior above $T_{\rm N}$, 
in particular near $T_{\rm S}$=260~K, 
we carried out weak transverse field (wTF-) $\mu^+$SR 
measurements up to 300~K. 
The wTF-$\mu^+$SR spectrum was fitted by 
a combination of a slowly and a fast relaxing precessing signal; 
the former is due to the external field and the latter due to the internal AF field 
(same as the first term in Eq.~(\ref{eq:ZFfit})); 
\begin{eqnarray}
 A_0 \, P_{\rm TF}(t) &=& 
A_{\rm TF}\cos(\omega_{\mu, \rm TF}t+\phi_{\rm TF})\exp(-\lambda_{\rm TF}t)~
\cr
&+& A_{\rm AF}\cos(\omega_{\mu, \rm AF}t+\phi_{\rm AF})\exp(-\lambda_{\rm AF}t), 
\label{eq:TFfit}
\end{eqnarray}
where 
$\omega_{\mu, \rm TF}$ and $\omega_{\mu, \rm AF}$ 
are the muon Larmor frequencies corresponding to 
the applied weak transverse field and the internal AF field, 
$\phi_{\rm TF}$ and $\phi_{\rm AF}$ are 
the initial phases of the two precessing signals and 
$A_n$ and $\lambda_n$ ($n$ = {\rm TF} and {\rm AF}) 
are the asymmetries and exponential relaxation rates of the two signals.  
Note that we have ignored the $J_0(\omega t)$ term
in Eq.~(\ref{eq:TFfit}) since
we are primarily interested in the magnetic behavior
above $T_{\rm N}$.

The results are shown in Fig.~$\ref{fig:wTF}$ together with $\chi^{-1}$. 
Besides the transition at 56~K, 
there are no anomalies up to 300~K in the normalized asymmetries, 
the relaxation rate ($\lambda_{\rm TF}$) or
the shift of the muon precession frequency ($\Delta\omega_{\mu, \rm TF}$).
Transverse field (TF-) $\mu^+$SR measurements 
at $H$=2600~Oe, which should be about 
50 times more sensitive to frequency shifts than the wTF measurements,
show no obvious changes in  
$\Delta\omega_{\mu, \rm TF}$ at $T_{\rm S}$ either.
On the other hand,
the $\chi^{-1}(T)$ curve exhibits a clear change in slope at $T_{\rm S}$. 
Above 60~K, the normalized wTF-asymmetry ($A_{\rm TF}$) levels off to
its maximum value 
--- i.e. the sample volume is almost 100\% paramagnetic.  
This therefore suggests that 
$T_{\rm S}$ is induced by a purely structural transition 
and there is no dramatic change in the spin state of Ni ions; 
that is, $T_{\rm S}$ is unlikely to be a cooperative JT transition. 
This is consistent with the fact that 
the crystal structure remains rhombohedral down to 5~K.\cite{A2NO_3}

\section{\label{sec:Su}Summary}

Positive muon spin rotation/relaxation ($\mu^+$SR) spectroscopy 
has been used to investigate the magnetic properties of 
a powder sample of Ag$_2$NiO$_2$ 
in the temperature range between 1.9 and 300~K.  
Zero field $\mu$SR measurements suggest the existence of 
an incommensurate antiferromagnetic (AF) order below $T_{\rm N}$=56~K.
An additional transition was also found at $T_{\rm m}$=22~K 
by both $\mu^+$SR and susceptibility measurements. 

The current results, when
compared to the results in AgNiO$_2$, 
indicate that 
magnetism in the half-filled {\sf 2DTL} of the NiO$_2$ plane 
is strongly affected by the interlayer distance. 
In other words, 
the ground state of the half-filled NiO$_2$ plane is 
not a ferromagnetic (FM) ordered state or
an FM spin-liquid or spin-glass, but is instead an AF frustrated system.  
The FM behavior in NaNiO$_2$ 
is therefore thought to be induced by 
a Jahn-Teller induced trigonal distortion.

\begin{acknowledgments}
This work was performed at TRIUMF. 
We thank S.R. Kreitzman, B. Hitti, D.J. Arseneau of TRIUMF 
for help with the $\mu^+$SR experiments. 
JHB is supported at UBC by CIAR, NSERC of Canada, 
and at TRIUMF by NRC of Canada.
KHC is supported by NSERC of Canada.
\end{acknowledgments}


\end{document}